\documentclass{amsart}
\usepackage{a4wide}
\usepackage{amssymb}
\usepackage{amsmath}
\usepackage{amsfonts}
\usepackage{amsthm}
\usepackage{graphicx}
\usepackage{epsfig}
\usepackage{longtable}
\usepackage{paralist}
\usepackage[usenames,dvipsnames]{color}
\newtheorem{thm}{Theorem}%[section]

\newtheorem{prop}[thm]{Proposition}
\newtheorem{conj}{Conjecture}

\theoremstyle{definition}
\newtheorem{defn}[thm]{Definition}

\theoremstyle{remark}
\newtheorem*{rmk}{Remark}

\newcommand{\eps}{\varepsilon}

\newcommand{\DEF}{{:=}}
\newcommand{\FED}{{=:}}

\newcommand{\Cset}{\mathbb{C}}

\newcommand{\IL}{\mathbb{L}}

\newcommand{\PT}[1]{\mathbf{#1}}

\newcommand{\re}{\mathop{\mathrm{Re}}}

\DeclareMathOperator{\dd}{\mathrm{d}}
\DeclareMathOperator{\xctint}{I}
\DeclareMathOperator{\numint}{Q}

\DeclareMathOperator{\capDISCR}{{D_{\mathrm{C}}}}

\DeclareMathOperator{\gammafcn}{\Gamma}

\DeclareMathOperator{\DirichletL}{L}

\DeclareMathOperator{\ELLtwo}{{\mathbb{L}_2}}
\DeclareMathOperator{\ELLtwoDISCR}{D_{\mathrm{C}}^{\ELLtwo}}

\DeclareMathOperator{\zetafcn}{\zeta}

  \definecolor{dgreen}{RGB}{68,138,103}%{0,112,27}
  \definecolor{dblue2}{RGB}{0,85,127}

\allowdisplaybreaks[1]

\title{Optimal Discrete Riesz Energy and Discrepancy}
\author{J. S. Brauchart} %\textasteriskcentered
\thanks{\noindent The author is recipient of an {\sc APART}-fellowship of the Austrian Academy of Sciences at UNSW (2010).}

\date{\today}

\begin{document}

\address{J. S. Brauchart:
School of Mathematics and Statistics, 
University of New South Wales, 
Sydney, NSW, 2052, 
Australia }
\email{j.brauchart@unsw.edu.au}

\begin{abstract}
The Riesz $s$-energy of an $N$-point configuration in the Euclidean space $\mathbb{R}^{p}$ is defined as the sum of reciprocal $s$-powers of all mutual distances in this system. In the limit $s\to0$ the Riesz $s$-potential $1/r^s$ ($r$ the Euclidean distance) governing the point interaction is replaced with the logarithmic potential $\log(1/r)$. In particular, we present a conjecture for the leading term of the asymptotic expansion of the optimal $\IL_2$-discrepancy with respect to spherical caps on the unit sphere in $\mathbb{R}^{d+1}$ which follows from Stolarsky's invariance principle [Proc. Amer. Math. Soc. 41 (1973)] and the fundamental conjecture for the first two terms of the asymptotic expansion of the optimal Riesz $s$-energy of $N$ points as $N \to \infty$. %\MARKED{OliveGreen}{We discuss how Riesz energy can be used to estimate the spherical cap discrepancy of point configurations (on the sphere) using Erd{\H o}s-{T}ur\'an type and LeVeque type discrepancy bounds. Further, we present a conjecture for the leading term of the asymptotic expansion of the optimal $\IL_2$-discrepancy with respect to spherical caps on the unit sphere in $\mathbb{R}^{d+1}$.} 
\end{abstract}

 %; Secondary 11K36}

\maketitle

% \nocite{AbSt1992}

% \section{Introduction}

% \section{Introduction}

\section{The discrete Riesz $s$-energy problem}

The {\em Riesz $s$-energy} (real $s\neq 0$) and the {\em logarithmic energy} ($s = 0$, by convention) of an $N$-point configuration $X_N$ with points $\PT{x}_1, \dots, \PT{x}_N$ in the Euclidean space $\mathbb{R}^p$ ($p \geq 1$) 
% modeling a discrete charge distribution of positive unit charges 
are defined as %the sums
\begin{equation*}
E_s(X_N) \DEF \mathop{\sum_{j=1}^N \sum_{k=1}^N}_{j \neq k} \frac{1}{\left| \PT{x}_j - \PT{x}_k \right|^s}, \qquad E_0(X_N) \DEF \mathop{\sum_{j=1}^N \sum_{k=1}^N}_{j \neq k} \log \frac{1}{\left| \PT{x}_j - \PT{x}_k \right|}. 
\end{equation*}
The logarithmic energy can be understood as the limiting case $s \to 0$. For large $s$ the nearest neighbor's interaction dominates and the other limiting case $s \to \infty$ gives the {\em best-packing problem} (or {\em Tammes problem} \cite{tammes:1930}). Most deeply studied is the classical ({\em Newtonian}) case $s = p - 2$ for the harmonic potential $1/r^{p-2}$, naturally with an abundance of literature in physics in the Coulomb case ($p = 3$), where the Riesz $s$-energy essentially is the potential energy of an ensemble of (positive) unit point charges placed at the points of the configuration $X_N$. 

For this note of interest are {\em sums of distances} (that is $s = -1$) for points on the sphere $\mathbb{S}^d$ because of the close connection to the spherical cap $\mathbb{L}_2$-discrepancy defined in Definition~\ref{def:sph.cap.L.2.discr} by means of Stolarsky's invariance principle (Proposition~\ref{prop:Stolarsky.inv.principle}) and to the worst-case error for equal weight quadrature formula for functions in the unit ball in a certain reproducing kernel Hilbert space over $\mathbb{S}^d$ (cf. \cite{BrDi2011_pre} and \cite{BrWo20xx}).

The {\em $N$-point $s$-energy} of an infinite compact set $A \subseteq \mathbb{R}^p$ is defined as follows:
\begin{align*}
\mathcal{E}_s( A; N ) &\DEF \sup \left\{ E_s( \PT{x}_1, \dots, \PT{x}_N ) : \PT{x}_1, \dots, \PT{x}_N \in A \right\}  &\text{if $s < 0$,} \\
\mathcal{E}_s( A; N ) &\DEF \inf \left\{ E_s( \PT{x}_1, \dots, \PT{x}_N ) : \PT{x}_1, \dots, \PT{x}_N \in A  \right\} &\text{if $s \geq 0$.}
\end{align*}
It gives the optimal $s$-energy an $N$-point configuration $X_N$ in $A$ can assume; that is 
\begin{equation*}
\mathcal{E}_s( A; N ) = E_s(X_N^{(s)}) \qquad \text{for an {\em optimal $s$-energy system $X_N^{(s)}$}.}
\end{equation*}
Let $d$ denote the Hausdorff dimension of $A$. When normalized appropriately (such that the total 'charge' of the $N$-point configuration is one) the quantity $\mathcal{E}_s( A; N ) / N^2$ remains finite (and has a limit as $N \to \infty$)\footnote{Indeed, for example, for $s > 0$ the sequence $N(N-1)/\mathcal{E}_s( A; N )$ is monotone and bounded and converges to the 'generalized transfinite diameter of $A$ of order $s$' defined by P{\'o}lya and  Szeg{\H{o}}~\cite{PoSz1931}.} in the {\em potential-theoretical regime $s < d$} but grows beyond any bound in the {\em hypersingular case} $s \geq d$. In the regime $s<d$ tools and result from classical potential theory (see, for example, Bj{\"o}rck~\cite{Bj1956} ($s<0$), Saff and Totik~\cite{SaTo1997} (logarithmic case) and Landkof~\cite{La1972}) are used to treat the limit $N\to\infty$. The limit distribution of optimal $s$-energy configurations is given by the (unique if $-2<s<d$) equilibrium (or extremal) measure $\mu_A$ maximizing (if $s<0$) or minimizing (if $s\geq0$) the energy integral 
\begin{equation*}
\mathcal{I}_s[\mu] \DEF \int \int k_s(\PT{x}, \PT{y}) \dd \mu(\PT{x}) \dd \mu(\PT{y}),
\end{equation*}
where $k_s(\PT{x}, \PT{y}) \DEF | \PT{x} - \PT{y} |^{-s}$ if $s\neq0$ and $k_0(\PT{x}, \PT{y}) \DEF - \log | \PT{x} - \PT{y} |$, among all Borel probability measures $\mu$ supported on $A$. The extremal quantity $V_s(A) = \mathcal{I}_s[\mu_A]$ is called the {\em $s$-energy of the set $A$}. If $s \geq d$, then $\mathcal{I}_s[\mu] = + \infty$ for any measure $\mu$ supported on $A$, that is $V_s(A) = +\infty$. In this case methods and results from geometrical measure theory are successfully used to gain more insight. One outcome is that other normalizations of the optimal $s$-energy would give a converging sequence for certain classes of compact sets (see, for example, Kuijlaars and Saff~\cite{KuSa1998} and Hardin and Saff~\cite{HaSa2005}). These results are the building blocks for the fundamental conjecture below.

The discrete Riesz energy problem is concerned with investigating properties of sequences of optimal (and nearly optimal) Riesz $s$-energy configurations. Questions concern 
\begin{inparaenum}[\itshape (i)]
\item explicit computations of optimal configurations,
\item limit distribution,
\item asymptotic expansion of the optimal energy (cf. Smale's Problem \#7~\cite{Sm1998}),
\item geometric properties ('well-separation', 'mesh'-norm).
\end{inparaenum}
% to name but a few.
%
We refer the interested reader to the survey articles Saff and Kuijlaars~\cite{SaKu1997} and Hardin and Saff~\cite{HaSa2004}.% regarding the discrete Riesz energy problem.

% This singles out the boundary point $s = d$ a turning point when 

% Let $\mathbb{S}^d$ denote the unit sphere in the Euclidean space $\mathbb{R}^{d+1}$ provided with the usual inner product $\langle \PT{\cdot}, \PT{\cdot} \rangle$. 
Let $\mathbb{S}^d$ denote the unit sphere in $\mathbb{R}^{d+1}$. The following long-standing open fundamental conjecture for the asymptotic expansion of the optimal Riesz $s$-energy is known. (In the upcoming paper \cite{BrHaSa2009a_pre} this conjecture and its history will be discussed in more detail.) Let $\mathcal{H}_d$ denote the $d$-dimensional Hausdorff measure (normalized so that the $d$-dimensional unit cube in $\mathbb{R}^p$ has measure $1$). 

\begin{conj}[Fundamental Conjecture] \label{conj:Riesz.s.energy.Sd}
Let $d \geq 2$ and $-2 < s < d + 2$ ($s \neq 0,d$).\footnote{For $s<-2$ the extremal distribution is concentrated in no more than $d+1$ points on $\mathbb{S}^d$ (\cite{Bj1956}). } Then 
\begin{equation*}
\mathcal{E}_s( \mathbb{S}^d; N ) = V_s( \mathbb{S}^d ) \, N^2 + \frac{C_{s,d}}{\left[ \mathcal{H}_d( \mathbb{S}^d ) \right]^{s/d}} \, N^{1 + s / d} + \mathcal{R}_s( \mathbb{S}^d; N ),
\end{equation*}
where $\mathcal{R}_s( \mathbb{S}^d; N ) / N^{1 + s / d - \eps} \to 0$ as $N \to \infty$ for some $\eps > 0$ possibly depending on $d$ and $s$.
\end{conj}

This conjecture combines known results on the basis of the {\em principle of analytic continuation}:
\begin{itemize}
\item In the potential-theoretical regime $-2 < s < d$ ($s \neq 0$) the dominant term grows like $N^2$ and the coefficient is given by the $s$-energy of the unit sphere
\begin{equation} \label{eq:V.s.S.d}
V_s( \mathbb{S}^d ) = 2^{d-s-1} \frac{\gammafcn((d+1)/2) \gammafcn((d-s)/2)}{\sqrt{\pi}\gammafcn(d-s/2)}, \qquad -2 < s < d (s \neq 0).
\end{equation}
For other values of $s$ it is understood to be the analytic continuation of the right-hand side above to the complex $s$-plane except at the simple poles at $s = d, d + 2, \dots, 2 d - 2$ (finitely many) if $d$ is even and at $s = d, d + 2, d + 4, \dots$ (infinitely many) if $d$ is odd.
\item In the hyper-singular case $s > d$ the leading term behaves like $N^{1+s/d}$ and it is shown by Kuijlaars and Saff  \cite{KuSa1998} that the limit $\mathcal{E}_s( \mathbb{S}^d; N ) / N^{1+s/d}$ exists.\footnote{The existence of this limit is proven for the class of rectifiable $d$-dimensional manifolds in \cite{HaSa2005} and for weighted Riesz $s$-energy for rectifiable $d$-dimensional sets  in \cite{BoHaSa2008}.} It is believed that the constant $C_{s,d}$ can be analytically continued to the $s$-plane, which is supported by the conjecture for its value in case of $d=2$ also provided in \cite{KuSa1998}. 
\item In the boundary case $s=d$ \cite{KuSa1998} gives that
\begin{equation*}
\mathcal{E}_d( \mathbb{S}^d; N ) \sim \frac{\mathcal{H}_d(\mathbb{B}^d)}{\mathcal{H}_d(\mathbb{S}^d)} \, N^2 \log N \qquad \text{as $N \to \infty$,}
\end{equation*}
which can be also understood by means of a limit process $s \to d$ assuming that the poles of $V_s(\mathbb{S}^d)$ and $C_{s,d}$ at $s=d$ cancel each other. (Obviously, one can also consider limit processes $s \to s^\prime$, where $s^\prime$ is a pole of $V_s$. In such a case one would gain information about the singularity at $s=s^\prime$ of the coefficient of the corresponding term in the asymptotic expansion of the optimal $s$-energy ($s$ near $s^\prime$) provided this lesser-order term exists. The limit process $s\to0$ connects the asymptotic expansion of the optimal logarithmic energy and optimal $s$-energy for $s$ near $0$.)
\end{itemize}
The constant $\mathcal{H}_d(\mathbb{B}^d) / \mathcal{H}_d(\mathbb{S}^d)$ is the ratio of the volume of the unit ball $\mathbb{B}^d$ in $\mathbb{R}^d$ and the surface area of $\mathbb{S}^d$ (denoted by $\omega_d$) and can be expressed in terms of a ratio of surface areas or a ratio of gamma functions
\begin{equation} \label{eq:ratio}
\frac{\mathcal{H}_d(\mathbb{B}^d)}{\mathcal{H}_d(\mathbb{S}^d)} = \frac{1}{d} \frac{\omega_{d-1}}{\omega_d} = \frac{1}{d} \, \frac{\gammafcn((d+1)/2)}{\sqrt{\pi} \gammafcn(d/2)} \sim \frac{1}{\sqrt{2\pi}} \, d^{-1/2} \quad \text{(as $d \to \infty$).}
\end{equation}

% Conjecture~\ref{conj:Riesz.sphere} combines known results described below on the basis of the {\em principle of analytic continuation}. 
% 
% \begin{conj} \label{conj:Riesz.2.sphere}
% For $-2 < \re s < 4$, $s \neq 2$, there holds
% \begin{equation*}
% \mathcal{E}_s( \mathbb{S}^2; N ) = \frac{2^{1-s}}{2-s} \, N^2 + \frac{\left( \sqrt{3} / 2 \right)^{s/2} \zetafcn_{\Lambda}(s)}{\left( 4 \pi \right)^{s/2}} \, N^{1+s/2} + \mathcal{O}_s( N^{-1+s/2}) \qquad \text{as $N \to \infty$.}
% \end{equation*}
% (Note that here $s=0$ does not refer to the logarithmic case. In fact, $\mathcal{E}_0( \mathbb{S}^2; N ) = N ( N - 1 )$ and any $N$-point configuration would be optimal in this case.)
% % where $C_{s,2} = \left( \sqrt{3} / 2 \right)^{s/2} \zetafcn_{\Lambda}(s)$. % The $\mathcal{O}$-constant remains bounded as $s\to0$.
% \end{conj}

It is a long-standing open problem what the precise value of the constant 
\begin{equation} \label{eq:C.s.d}
C_{s,d} = \lim_{N \to \infty} \mathcal{E}_s( [0,1]^d; N ) \big/ N^{1+s/d}
\end{equation}
is for $d \geq 2$. (In the case $d = 1$ Mart{\'{\i}}nez-Finkelshtein et al. showed that $C_{s,1}=2 \zeta(s)$, where $\zeta(s)$ is the Riemann zeta function.) The significance of the constant $C_{s,d}$ follows from the fact that the limit $\lim_{s\to\infty}[ C_{s,d} ]^{1/s}$ can be expressed in terms of the 'largest sphere packing density' in $\mathbb{R}^d$, which is only known in three cases: $d = 1,2$ and $d = 3$ (Kepler conjecture, rather recently proved by Hales~\cite{Ha2005}). For $d = 2$ Kuijlaars and Saff \cite{KuSa1998} obtained the estimate
\begin{equation*} 
\limsup_{N \to \infty} \frac{\mathcal{E}_s( \mathbb{S}^2; N )}{N^{1+s/2}} \leq \frac{\left( \sqrt{3} / 2 \right)^{s/2} \zetafcn_{\Lambda}(s)}{\left( 4 \pi \right)^{s/2}}, \qquad s > 2,
\end{equation*}
and they conjecture that equality holds above; that is:
\begin{conj}[Kuijlaars and Saff \cite{KuSa1998}] \label{conj:C.s.2}
For $s > 2$, $C_{s,2} = \left( \sqrt{3} / 2 \right)^{s/2} \zetafcn_{\Lambda}(s)$.
\end{conj}

The function $\zetafcn_\Lambda(s)$ is the zeta function of the hexagonal lattice $\Lambda = \{ m \, (1, 0) + n (1/2, \sqrt{3}/2) : m, n \in \mathbb{Z} \}$, which was used in \cite{KuSa1998} to locally approximate a minimal $s$-energy configuration on the sphere in the hypersingular case $s>2$. For $\re s > 2$ it is defined as
\begin{equation*}
\zetafcn_\Lambda(s) = \sum_{\PT{0} \neq \PT{a} \in \Lambda} \frac{1}{\left| \PT{a} \right|^s} = \sum_{(m,n) \in \mathbb{Z}^2\setminus \{ (0,0) \}} \frac{1}{\left( m^2 + m n + n^2 \right)^{s/2}}
\end{equation*}
and it is understood that $\zetafcn_\Lambda(s)$ is the meromorphic extension to $\mathbb{C}$ of the right-hand side above.
The function $\zetafcn_\Lambda(s)$ admits a factorization (cf., for example, \cite[Chapter~X, Section~7]{Co1980})
\begin{equation} \label{zeta.lambda.prod}
\zetafcn_\Lambda(s) = 6 \zetafcn( s / 2 ) \DirichletL_{-3}( s / 2 ), \qquad \re s > 2,
\end{equation}
into a product of the Riemann zeta function $\zetafcn$ and the first negative primitive Dirichlet $\DirichletL$-Series
\begin{equation} 
\DirichletL_{-3}(s) \DEF 1 - \frac{1}{2^s} + \frac{1}{4^s} - \frac{1}{5^s} + \frac{1}{7^s} - \cdots, \qquad \re s > 1.
\end{equation}
Interestingly, it is assumed in physics that a hexagonal configuration has lowest potential energy but there seems to be no mathematical proof for this. However, it is known that the hexagonal lattice is (even universal) optimal among all lattices in $\mathbb{R}^2$, see Montgomery~\cite{Mo1988} and Cohn and Kumar~\cite{CoKu2007}.

\section{Uniform Distribution and Discrepancy}

\begin{defn} \label{def:asymptotically.uniformly.distributed}
A sequence $\{ X_N \}$ is {\em asymptotically uniformly distributed on $\mathbb{S}^d$} if
\begin{equation*} 
\lim_{N\to\infty} \frac{\#\left\{ k : \PT{x}_{k,N} \in B \right\}}{N} = \sigma_d(B)
\end{equation*}
for every $\sigma_d$-measurable clopen set $B$ in $\mathbb{S}^d$.
\end{defn}

Informally speaking: a reasonable test set gets a fair share of points as $N$ becomes large.

The quality of a sequence $\{ X_N \}$ of $N$-point systems can be quantified using the discrepancy
\begin{equation*}
D(\mathcal{F};\PT{x}_1,\dots,\PT{x}_N) \DEF \sup_{B\in\mathcal{F}} \left| \frac{\#\left\{ k : \PT{x}_k\in B \right\}}{N} - \sigma_d(B) \right| 
\end{equation*}
measuring the maximum deviation between the uniform measure (limit distribution of optimal configurations) and empirical point distribution with respect to a family $\mathcal{F}$ of test sets (for example spherical caps). The {\em spherical cap discrepancy} of an $N$-point configuration $X_N$ will be denoted by $\capDISCR( X_N )$.

A different approach to uniform distribution makes use of numerical integration and function spaces. Let 
\begin{equation*}
\xctint[f] \DEF \int_{\mathbb{S}^d} f \dd \sigma_d, \qquad \displaystyle Q_N[f] \DEF \frac{1}{N} \sum_{k=1}^N f(\PT{x}_k).
\end{equation*}

\begin{defn} \label{def:equi-distributed.cont.fcns}
$\{ X_N \}$ is {\em equi-distributed with respect to every function in the space of continuous functions} if \footnote{ That is, the discrete probability measure associated with $X_N$ tends to $\sigma_d$ as $N \to \infty$ (in the weak-$*$ limit). }
\begin{equation*}
\lim_{N \to \infty} \numint_N[f] = \xctint[f] \qquad \text{for every $f \in C(\mathbb{S}^d)$.}
\end{equation*}
\end{defn}
As it is well-known both Definitions \ref{def:asymptotically.uniformly.distributed} and \ref{def:equi-distributed.cont.fcns} are equivalent. 

On the sphere one has no satisfactory analogue to the celebrated Koksma-Hlawka inequality in the unit cube. Cui and Freeden~\cite{CuFr1997} introduced the concept of {\em generalized discrepancy on $\mathbb{S}^2$} based on pseudo-differential operators. For a particular choice $\mathbf{D}$ they obtained a {\em Koksma-Hlawka} like inequality 
\begin{equation*}
\big| \numint_N[f] - \xctint[f] \big| \leq \sqrt{6} \, \mathrm{D}_{\mathrm{CF}}(X_N) \, \| f \|_{\mathbb{H}^{3/2}(\mathbb{S}^2)},
\end{equation*}
where $f$ is from a certain Sobolev space $\mathbb{H}^{3/2}(\mathbb{S}^2)$ whose reproducing kernel is defined using $\mathbf{D}$. In this context, a sequence $\{X_N\}$ of $N$-point systems is called {\em $\mathbf{D}$-equidistributed with respect to all functions in $\mathbb{H}^{3/2}(\mathbb{S}^2)$} if $\lim_{N\to\infty} \mathrm{D}_{\mathrm{CF}}(X_N) = 0$. Moreover, the generalized discrepancy associated with $\mathbf{D}$ has a closed form expressible in terms of elementary functions
\begin{equation*}
4 \pi \left[ \mathrm{D}_{\mathrm{CF}}(X_N) \right]^2 = 1 - \frac{1}{N^2} \sum_{j, k = 1}^N \log \left( 1 + \left| \PT{x}_j - \PT{x}_k \right| / 2 \right)^2.
\end{equation*}
(Sloan and Womersley~\cite{SlWo2004} showed that $\left[ \mathrm{D}_{\mathrm{CF}}(X_N) \right]^2$ has a natural interpretation as the worst-case error for $Q_N$ for function from the unit ball in $\mathbb{H}^{3/2}(\mathbb{S}^2)$ provided with a norm which is equivalent to the one used by Cui and Freeden.) This approach is followed further in \cite{BrWo20xx} leading to the generalized discrepancy
\begin{equation} \label{eq:sum.dist.discr}
\left[ \mathrm{D}(X_N) \right]^2 = \frac{4}{3} - \frac{1}{N^2} \sum_{k, \ell = 1}^N \left| \PT{x}_\ell - \PT{x}_k \right|
\end{equation}
associated with the Sobolev space $\mathbb{H}^{3/2}(\mathbb{S}^2)$ with the reproducing kernel $K(\PT{x}, \PT{y}) = (8 / 3) - |\PT{x} - \PT{y} |$.

\section{The spherical cap $\ELLtwo$-discrepancy on $\mathbb{S}^d$}

A spherical cap on $\mathbb{S}^d$ centered at $\PT{x} \in \mathbb{S}^d$ is the set 
\begin{equation*}
C(\PT{x}; t) \DEF \left\{ \PT{y} \in \mathbb{S}^d : \left\langle \PT{x}, \PT{y} \right\rangle \geq t \right\},
\end{equation*}
where $\langle \PT{\cdot}, \PT{\cdot} \rangle$ denotes the usual inner product. The family of all spherical caps on $\mathbb{S}^d$ forms a {\em discrepancy system} (see \cite{DrTi1997}) giving rise to the notion of {\em asymptotic uniform distribution}

\begin{defn} \label{def:sph.cap.L.2.discr}
The {\em spherical cap $\ELLtwo$-discrepancy} of an $N$-point configuration $X_N$ {on $\mathbb{S}^d$} is given by
\begin{equation*}
\ELLtwoDISCR( X_N ) \DEF \left[ \int_{-1}^1 \int_{\mathbb{S}^d} \left| \frac{\left| X_N \cap C( \PT{x}, t) \right|}{N} - \sigma_d( C( \PT{x}, t ) ) \right|^2 \dd \sigma_d( \PT{x} ) \dd t \right]^{1/2}.
\end{equation*}
The {\em optimal $\ELLtwo$-discrepancy} of $N$-point configurations on $\mathbb{S}^d$ is denoted by
\begin{equation*}
\ELLtwoDISCR( \mathbb{S}^d; N ) \DEF \inf \left\{ \ELLtwoDISCR( X_N ) : X_N \subseteq \mathbb{S}^d \right\}.
\end{equation*}
\end{defn}

An obvious upper bound of the $\ELLtwo$-discrepancy is in terms of the spherical cap discrepancy: 
% For any $N$-point system $X_N$ on $\mathbb{S}^d$ 
\begin{equation*}
\ELLtwoDISCR( X_N ) \leq \sqrt{2} \, \capDISCR( X_N ), \qquad X_N \subseteq \mathbb{S}^d.
\end{equation*}

% {\em Stolarsky's invariance principle} \cite{St1973} \footnote{In \cite{Br2003} we extend this principle.} connects the sum of distances, the $\ELLtwo$-discrepancy and the so-called distance integral.
The next result connects the sum of distances, the $\ELLtwo$-discrepancy and the distance integral.
\begin{prop}[Stolarsky's invariance principle~\cite{St1973}] \label{prop:Stolarsky.inv.principle} \footnote{The given version is a special case of the result in \cite{St1973} which is extended in \cite{Br2003}.} Let $d \geq 2$. Then %for any $N$-point configuration $X_N$ on $\mathbb{S}^d$: 
\begin{equation*}
\frac{1}{N^2} \sum_{j, k = 1}^N \left| \PT{x}_j - \PT{x}_k \right| + \frac{\mathcal{H}_d(\mathbb{S}^d)}{\mathcal{H}_d(\mathbb{B}^d)} \left[ \ELLtwoDISCR( X_N ) \right]^2 = \int \int \left| \PT{x} - \PT{x}^\prime \right| \dd \sigma_d(\PT{x}) \dd \sigma_d(\PT{x}^\prime), \qquad X_N \subseteq \mathbb{S}^d.
\end{equation*}
\end{prop}
Clearly, reduction of the $\ELLtwo$-discrepancy means increase of the sum of distances and vice versa. 
% Recall, that the reciprocal of the constant above also appears in the leading term of the asymptotic expansion of the minimal Riesz $d$-energy.

Stolarsky's invariance principle provides a simple way to compute the $\ELLtwo$-discrepancy of a given $N$-point configuration on $\mathbb{S}^d$. In the language of the discrete energy problem we have 
\begin{equation} \label{eq:ELLtwoDISCR}
\frac{\mathcal{H}_d(\mathbb{S}^d)}{\mathcal{H}_d(\mathbb{B}^d)} \left[ \ELLtwoDISCR( X_N ) \right]^2 = V_{-1}(\mathbb{S}^d) - \frac{1}{N^2} \sum_{j, k = 1}^N \left| \PT{x}_j - \PT{x}_k \right| = - \left[ E_{-1}(X_N) - V_{-1}(\mathbb{S}^d) \, N^2 \right] / N^2. % = V_{-1}(\mathbb{S}^d) - \frac{E_{-1}(X_N)}{N^2} .
\end{equation}
Evidently, the square of the optimal $\ELLtwo$-disrepancy is closely related to the second term of the asymptotic expansion of the optimal Riesz $(-1)$-energy, cf. Conjecture~\ref{conj:Riesz.s.energy.Sd}. This connection between the spherical cap ($\mathbb{L}_2$-)discrepancy and the second term in the asymptotic expansion of the $s$-energy is exploited in \cite{Br2008}, Grabner and Damelin~\cite{DaGr2003}.
% Recall, that the middle part in \eqref{eq:ELLtwoDISCR} has a natural interpretation as the square of the worst-case error for functions in the unit ball in the Sobolev space $\mathbb{H}^{s}(\mathbb{S}^d)$ with $s = (d + 1) / 2$ provided with a certain reproducing kernel and can be seen as a discrepancy in the sense of Definition~\ref{def:}, see Section~\ref{sec:}.

The $\ELLtwo$-discrepancy of an $N$-point configuration $X_N$ is minimial if and only if the sum of distances of $X_N$ is maximal. Thus, optimal $\ELLtwo$-discrepancy configurations are, in fact, maximal sum-of-distance configurations or, equivalently, maximal Riesz $(-1)$-energy configurations. Such configurations can be generated using numerical optimization which is a highly non-linear process.  

Stolarsky used his invariance principle and discrepancy results of Schmidt~\cite{Sch1969} on the discrepancy of spherical caps to estimate the difference
\begin{equation*}
\int \int \left| \PT{x} - \PT{x}^\prime \right| \dd \sigma_d(\PT{x}) \dd \sigma_d(\PT{x}^\prime) - \frac{\mathcal{E}_{-1}( \mathbb{S}^d; N )}{N^2}
\end{equation*}
and obtained the correct order of $N$ for the upper bound. Harman~\cite{Ha1982} improved the lower bound (for the general invariance principle) and, finally, Beck~\cite{Be1984} obtained the correct order of $N$ for the Euclidean metric. By means of the invariance principle these bound for the energy difference translate into bounds for the spherical cap $\ELLtwo$-discrepancy.
% From Beck's result concerning lower bound and Stolarsky's result concerning upper bounds of the middle part in \eqref{eq:ELLtwoDISCR}, see relation~\eqref{eq:Beck}, there follows lower and upper bounds for the $\ELLtwo$-discrepancy.
% 
\begin{prop} \label{prop:ELLtwoDISCR.bounds}
Let $d \geq 2$. There exist constants $c^\prime, C^\prime > 0$ such that for sufficiently large $N$
\begin{equation*}
c^\prime \, N^{-1/2 - 1/(2d)} \leq \ELLtwoDISCR( \mathbb{S}^d; N ) \leq C^\prime \, N^{-1/2 - 1/(2d)}.
\end{equation*}
\end{prop}

\begin{rmk}
By Proposition~\ref{prop:ELLtwoDISCR.bounds} the correct order of the decay of the optimal $\ELLtwo$-discrepancy on $\mathbb{S}^d$ in terms of powers of the number of points $N$ is $N^{-1/2 - 1/(2d)}$ which is the same rate (apart from the $\sqrt{\log N}$ term in the upper bound) as for the optimal spherical discrepancy on $\mathbb{S}^d$.
\end{rmk}

Proposition~\ref{prop:ELLtwoDISCR.bounds} rises the question if $\ELLtwoDISCR( \mathbb{S}^d; N ) \, N^{1/2 + 1/(2d)}$ has a limit as $N \to \infty$. By employing the connection to the discrete Riesz $s$-energy problem, where one has conjectures regarding the asymptotic expansion of the optimal $s$-energy, we derive the following conjectures for the leading term in the asymptotic expansion of the optimal $\ELLtwo$-discrepancy.
% \begin{conj} \label{conj:L2discr.S2}
% Let $\{ X_N^* \}_{N \geq2}$ be a sequence of $N$-point configurations on $\mathbb{S}^2$ with optimal $\IL_2$-discrepancy $\ELLtwoDISCR( X_N^* )$ for every $N$. If Conjecture~\ref{conj:Riesz.s.energy.S2} holds, then 
% \begin{equation*}
% \ELLtwoDISCR( X_N^* ) \sim (?) N^{-1/2 - 1/(2d)} + \cdots \qquad \text{as $N \to \infty$.}
% \end{equation*}
% \end{conj}
\begin{conj} \label{conj:L2discr.Sd}
Let $d \geq 2$. If Conjecture~\ref{conj:Riesz.s.energy.Sd} holds, then 
\begin{equation} \label{eq:A.d}
\ELLtwoDISCR( \mathbb{S}^d; N ) \sim A_d \, N^{-1/2 - 1/(2d)} + \cdots \qquad \text{as $N \to \infty$,} \qquad A_d \DEF \sqrt{\frac{\mathcal{H}_d(\mathbb{B}^d)}{\mathcal{H}_d(\mathbb{S}^d)} \, \frac{-C_{-1,d}}{\left[ \mathcal{H}_d( \mathbb{S}^d ) \right]^{-1/d}}}.
\end{equation}
% where
% \begin{equation} \label{eq:A.d}
% A_d \DEF \sqrt{\frac{\mathcal{H}_d(\mathbb{B}^d)}{\mathcal{H}_d(\mathbb{S}^d)} \, \frac{-C_{-1,d}}{\left[ \mathcal{H}_d( \mathbb{S}^d ) \right]^{-1/d}}}.
% \end{equation}
\end{conj}

\begin{proof}[Justification]
By Stolarsky's invariance principle (Prop.~\ref{prop:Stolarsky.inv.principle} and Eq.~\eqref{eq:ELLtwoDISCR})
\begin{equation*}
\frac{\mathcal{H}_d(\mathbb{S}^d)}{\mathcal{H}_d(\mathbb{B}^d)} \left[ \ELLtwoDISCR( \mathbb{S}^d; N ) \right]^2 = - \left[ \mathcal{E}_{-1}(\mathbb{S}^d; N) - V_{-1}( \mathbb{S}^d ) \, N^2 \right] / N^2.
\end{equation*}
Application of the fundamental Conjecture~\ref{conj:Riesz.s.energy.Sd} yields
\begin{equation*}
\frac{\mathcal{H}_d(\mathbb{S}^d)}{\mathcal{H}_d(\mathbb{B}^d)} \left[ \ELLtwoDISCR( \mathbb{S}^d; N ) \right]^2 = \frac{-C_{-1,d}}{\left[ \mathcal{H}_d( \mathbb{S}^d ) \right]^{-1/d}} \, N^{-1 - 1 / d} - \mathcal{R}_{-1}( \mathbb{S}^d; N ) / N^2.
\end{equation*}
The conjecture follows.
\end{proof}

For the unit sphere $\mathbb{S}^2$ one can make the conjecture more precise.

\begin{conj} \label{conj:L2discr.S2}
If Conjectures~\ref{conj:Riesz.s.energy.Sd} and \ref{conj:C.s.2} hold, then 
\begin{equation*}
\ELLtwoDISCR( \mathbb{S}^2; N ) \sim A_2 \, N^{-1/2 - 1/(2d)} + \cdots \qquad \text{as $N \to \infty$,}
\end{equation*}
where 
\begin{equation*}
A_2 
% = \sqrt{ \frac{ \left( 3 / 4 \right)^{3/4} }{ \sqrt{\pi} } \zetafcn(3/2) \DirichletL_{-3}(-1/2) } 
= \sqrt{ \frac{3}{2} \left( \frac{8 \pi}{\sqrt{3}} \right)^{1/2} \left[ - \zetafcn(-1/2) \right] \DirichletL_{-3}(-1/2) }
= 0.44679728350408\dots
% \sqrt{\left( 3 / 2 \right) \zeta(3/2) \DirichletL_{-3}(-1/2)} = 0.84122658671\dots.
\end{equation*}
\end{conj}

\begin{proof}[Justification]
Since $\mathcal{H}_d( \mathbb{S}^d ) = \omega_2 = 4 \pi$ and (by \eqref{eq:ratio})
\begin{equation*}
\frac{\mathcal{H}_d(\mathbb{B}^d)}{\mathcal{H}_d(\mathbb{S}^d)} \Bigg|_{d=2} = \frac{1}{2} \, \frac{\gammafcn(3/2)}{\sqrt{\pi} \gammafcn(1)} = \frac{1}{4},
\end{equation*}
one obtains (using Conjecture~\ref{conj:C.s.2} and relations \eqref{zeta.lambda.prod} and \eqref{eq:A.d})
\begin{align*}
A_2 
&= \sqrt{\frac{1}{4} \, \frac{-C_{-1,2}}{\left( 4 \pi \right)^{-1/2}} } = \sqrt{\frac{1}{4} \left( \sqrt{3} / 2 \right)^{-1/2} \frac{-6 \zetafcn( -1 / 2 ) \DirichletL_{-3}( -1 / 2 )}{\left( 4 \pi \right)^{-1/2}} } \\
% &= \sqrt{\frac{3}{2} \left( \sqrt{3} / 2 \right)^{-1/2} \frac{-\zetafcn( -1 / 2 ) }{\left( 4 \pi \right)^{-1/2}} \, \DirichletL_{-3}( -1 / 2 ) } = \sqrt{\frac{3}{2} \left( \sqrt{3} / 2 \right)^{-1/2} \frac{\zetafcn( 3 / 2 ) }{\left( 4 \pi \right)^{1/2}} \, \DirichletL_{-3}( -1 / 2 ) } \\
% &= \sqrt{ \frac{ \left( 3 / 4 \right)^{3/4} }{ \sqrt{\pi} } \zetafcn(3/2) \DirichletL_{-3}(-1/2) } \\
&= \sqrt{ \frac{3}{2} \left( \frac{8 \pi}{\sqrt{3}} \right)^{1/2} \left[ - \zetafcn(-1/2) \right] \DirichletL_{-3}(-1/2) } = 0.44679728350408\dots
\end{align*}
\end{proof}

% In general, that is for higher-dimensional spheres, we propose:
% 
% \begin{conj} \label{conj:L2discr.Sd}
% Let $\{ X_N^* \}_{N \geq2}$ be a sequence of $N$-point configurations on $\mathbb{S}^d$ with optimal $\IL_2$-discrepancy $\ELLtwoDISCR( X_N^* )$ for every $N$. If Conjecture~\ref{conj:Riesz.s.energy.Sd} holds, then 
% \begin{equation*}
% \ELLtwoDISCR( X_N^* ) \sim \sqrt{-C_{-1,d} / 2} \left[ \mathcal{H}_d(\mathbb{S}^d) \right]^{-1/(2d)} N^{-1/2 - 1/(2d)} + \cdots \qquad \text{as $N \to \infty$.}
% \end{equation*}
% \end{conj}

% \section{The Worst-Case Error in a certain Sobolev Space}

\subsection*{The case $d=1$}

Stolarsky' invariance principle also holds for $d=1$ as one can see from the proof given in \cite{BrDi2011_pre}. In this case one has (cf. Eq.s~\eqref{eq:V.s.S.d} and \eqref{eq:ratio})
\begin{equation*}
\frac{1}{N^2} \sum_{j, k = 1}^N \left| \PT{x}_j - \PT{x}_k \right| + \frac{1}{\pi} \left[ \ELLtwoDISCR( X_N ) \right]^2 = \frac{4}{\pi}, \qquad X_N \subseteq \mathbb{S}^1.
\end{equation*}
On the other hand (in joint work with Hardin and Saff)~\cite{BrHaSa2009} we obtained a complete asymptotic expansion of the Riesz $s$-energy $\mathcal{L}_{s}(N)$ of the $N$th roots of unity which represent optimal $N$-point configurations for $s > - 2$. In general, for $s \in \mathbb{C}$ with $s \neq 0, 1, 3, \dots$
\begin{equation*} \label{eq:cal.L.s.N}
\mathcal{L}_s(N) = V_s( \mathbb{S} ) \, N^2 + \frac{2\zetafcn(s)}{(2\pi)^s} N^{1+s} + \frac{2}{(2\pi)^{s}} \sum_{n=1}^{p} \alpha_n(s) \zetafcn(s-2n) \, N^{1+s-2n} + \mathcal{O}( N^{-1+ \re s-2p} ).
\end{equation*}
The coefficients $\alpha_n(s)$, $n\geq0$, satisfy the generating function relation
\begin{equation*} \label{sinc.power.0}
\left( \frac{\sin \pi z}{\pi z} \right)^{-s} = \sum_{n=0}^\infty \alpha_n(s) z^{2n},
\quad |z|<1, \ s\in \Cset. %, \qquad \alpha_n(s) = \frac{(-1)^n B_{2n}^{(s)}(s/2)}{(2n)!} \left( 2 \pi \right)^{2n}.
\end{equation*}
It follows that
\begin{equation*}
\frac{1}{\pi} \left[ \ELLtwoDISCR( X_N ) \right]^2 = \frac{2 [ - \zetafcn(-1) ]}{(2 \pi)^{-1}} \, N^{-2} + \frac{2}{(2 \pi)^{-1}} \sum_{n=1}^{p-1} \alpha_n(-1) \zetafcn(-1-2n) \, N^{-2-2n} + \mathcal{O}(N^{-2-2p}),
\end{equation*}
where ($B_0=1$, $B_1=-1/2$, \dots{}  are the so-called {\em Bernoulli numbers} and $(-1)^{n+1} B_{2n}>0$)
\begin{equation*}
\alpha_n(-1) \zetafcn(-1-2n) = \frac{(-1)^{n+1} B_{2(n+1)}}{(2(n+1))!} \pi^{2n} < 0, \qquad n \geq 1.
\end{equation*}
In the last step it was used that the Riemann zeta function at negative integers can be expressed in terms of Bernoulli numbers.

\subsection*{Numeric Results}

Recently, in joint work with Josef Dick we investigated the properties of so-called digital nets (see Niederreiter~\cite{Ni1988} and Dick and Pillichshammer~\cite{DiPi2010}) lifted to the sphere by means of an area preserving map. Numerical experiments for Sobol' Sequences lifted to $\mathbb{S}^2$ in \cite{BrDi2011b_pre} suggest that the order of $N$ is optimal but the constant seems to be too large (in the range $[0.50,0.59]$ for $N = 2^m$, $m=1,\dots,20$). (Optimal $(-1)$-energy configurations are more difficult to come by and the number of points for which numerical optimization is feasible  and the result could be trusted to be the global maximum is very limited. Cf. \cite{BaBr2009,BeCa2009,BeClDu2004} and references cited therein.) % when taking into account the experience for other values of $s$.) 

\section{Estimating the spherical cap discrepancy}

The classical {\em Erd{\"o}s-Tur{\'a}n type inequality} (cf. Grabner~\cite{Gr1991}, also cf. Li and Vaaler~\cite{LiVa1999}) estimate the spherical cap discrepancy in terms of Weyl sums. Recently, by generalizing LeVeques result for the unit circle \cite{LeV1965}, Narcowich et al.~\cite{NaSuWa2010} obtained {\em LeVeque type inequalities} on the sphere:
\begin{equation*}
c_1 \left[ \sum_{\ell=1}^\infty a_\ell \sum_{m=1}^{Z(d,\ell)} \left| \frac{1}{N} \sum_{j=0}^{N-1} Y_{\ell,m}(\PT{x}_{j}) \right|^2 \right]^{1/2} \leq D_{\mathrm{C}}(X_{N}) \leq c_2 \left[ \sum_{\ell=1}^\infty b_\ell \sum_{m=1}^{Z(d,\ell)} \left| \frac{1}{N} \sum_{j=0}^{N-1} Y_{\ell,m}(\PT{x}_{j}) \right|^2 \right]^{1/(d+2)},
\end{equation*}
where $a_\ell \DEF \Gamma(\ell - 1/2) / \Gamma( \ell + d + 1/2) \asymp 1 / \ell^{d+1} \FED b_\ell$ for some positive constant $c_1$ and $c_2$ and $Z(d,n)$ denotes the number of linearly independent real spherical harmonics $Y_{\ell,m}$ of degree $\ell$. In Sun and Chen~\cite{SuCh2008} 'spherical basis functions' (as a counter part to radial basis function on spheres utilized in \cite{NaSuWa2010}) are used to investigate uniform distribution on spheres.

G. Wagner obtained ground-breaking results concerning estimates of the Riesz $s$-energy and explored connections between $s$-energy and discrepancy (see \cite{Wa1990, Wa1992b, Wa1992}). In fact, the proofs of both the Erd{\"o}s-Tur{\'a}n type inequality and the LeVeque type inequality can be modified yielding estimates of the spherical cap discrepancy in terms of $s$-energy. A different approach exploits the connection between the error of numerical integration for polynomials (cf. Damelin and Grabner~\cite{DaGr2003}) and, by utilizing a result of Andrievskii, Blatt and G{\"o}tz~\cite{AnBlGo1999} (see this author~\cite{Br2008}). 

We close this note by recalling that optimal Riesz $s$-energy configurations may not have optimal spherical cap discrepancy (essentially of order $\mathcal{O}(N^{-1/2-1/(2d)})$, see Beck~\cite{Be1984}) which is reflected in Korevaar's conjecture~\cite{Ko1996} claiming for the harmonic case $s=d-1$ that the spherical cap discrepancy is of order $\mathcal{O}(N^{-1/d})$. This was essentially proved by G{\"o}tz~\cite{Go2000}. He also showed that the bound is sharp for $d=2$. The conjecture is still open for other values of $s$ and $d$.

\vspace{10mm}
{\bf Acknowledgement:} The author is grateful to the School of Mathematics and Statistics at UNSW for their support.

\def\cprime{$'$} \def\polhk#1{\setbox0=\hbox{#1}{\ooalign{\hidewidth
  \lower1.5ex\hbox{`}\hidewidth\crcr\unhbox0}}}

\end{document}